\documentclass[10pt]{cai}

\graphicspath{{figs/}}

\begin{document}

\volumeheader{34}{0}
\begin{center}

    \title{Anomaly Detection Approach to Identify Early Cases in a Pandemic using Chest X-rays}
    \maketitle

  \thispagestyle{empty}

  \begin{tabular}{cc}
    Shehroz S. Khan\upstairs{\affilone,*}, Faraz Khoshbakhtian\upstairs{\affiltwo}, Ahmed Bilal Ashraf \upstairs{\affilthree}
  \\[0.25ex]
  {\small \upstairs{\affilone} KITE, Toronto Rehabilitation Institute, University Health Network} \\
  {\small \upstairs{\affiltwo} University of Toronto} \\
  {\small \upstairs{\affilthree} University of Manitoba} \\
  \end{tabular}
  
  \emails{
    \upstairs{*}corresponding\_shehroz.khan@uhn.ca
    }
  \vspace*{0.2in}
\end{center}
\begin{abstract}

The current COVID-19 pandemic is now getting contained, albeit at the cost of more than $2.3$ million human lives. A critical phase in any pandemic is the early detection of cases to develop preventive treatments and strategies. 
In the case of COVID-19, several studies have indicated that chest radiography images of the infected patients show characteristic abnormalities. However, at the onset of a given pandemic, such as COVID-19, there may not be sufficient data for the affected cases to train models for their robust detection. Hence, supervised classification is ill-posed for this problem because the time spent in collecting large amounts of data from infected persons could lead to the loss of human lives and delays in preventive interventions. Therefore, we formulate the problem of identifying early cases in a pandemic as an anomaly detection problem, in which the data for healthy patients is abundantly available, whereas no training data is present for the class of interest (COVID-19 in our case). To solve this problem, we present several unsupervised deep learning approaches, including convolutional and adversarially trained autoencoder. We tested two settings on a publicly available dataset (COVIDx) by training the model on chest X-rays from (i) only healthy adults,  and (ii) healthy and other non-COVID-19 pneumonia, and detected COVID-19 as an anomaly. After performing $3$-fold cross validation, we obtain a  ROC-AUC of $0.765$. These results are very encouraging and pave the way towards research for ensuring emergency preparedness in future pandemics, especially the ones that could be detected from chest X-rays.
\end{abstract}

\begin{keywords}{Keywords:}
Chest X-ray, Autoencoder, Anomaly Detection, Convolutional Neural Network, Adversarial Learning
\end{keywords}
\copyrightnotice

\section{Introduction}
\label{sec:intro}
Detecting early cases in any pandemic is crucial, so that preventive measures could be taken. However, in the early phase of a pandemic, sufficient representative cases of the disease may not be available. Therefore, training machine learning models on such a skewed dataset is very challenging \cite{khan2014one}. In the case of COVID-19, even though the earliest cases of pandemic were reported in the media by the end of 2019 \cite{nytimes2020}, data availability was extremely sparse in the beginning, precluding the identification of pandemic onset and development of preventive measures. In the months that followed, it led to the loss of over $2.3$ million lives (by February 2021) \cite{who2021}. As of now, there are several COVID-19 vaccines available in the market \cite{raps2021} with promising results and the pandemic could be contained. Several lessons can be learnt from this unfortunate situation to avert millions of humans deaths in the future. One of the most important lessons is to be prepared for early detection of cases in a future pandemic, and we take COVID-19 as a use case in this paper.


COVID-19 is caused by the severe acute respiratory syndrome coronavirus 2 (SARS-CoV-2). Currently, reverse transcription-polymerase chain reaction is the gold standard screening method for this viral infection \cite{wang2020detection}. This test 
shows high specificity \cite{corman2020detection,lan2020positive}, albeit with variability and inconsistency \cite{west2020covid}, along with being expensive and very time-consuming. 
Several studies indicated promising results for screening COVID-19 patients by chest radiology imaging, including chest x-rays and CT scans \cite{huang2020clinical,luz2020towards,oh2020deep,wang2020covid,tuluptceva2020anomaly,fang2020sensitivity,gozes2020rapid}. While CT scans offer superior quality and 3D details of imaging \cite{fang2020sensitivity,gozes2020rapid}, they are costly,  require sanitation of the scanner after each scan, and may not be readily available in the healthcare ecosystems of low-income countries. On the other hand, Chest X-ray (CXR) is a promising imaging modality as it is easily available, and can be used for rapid triaging \cite{wang2020covid}. Despite the easy availability of an imaging modality, identifying early cases of COVID-19 remains a challenge. 
    
Most of the studies that use CT-scans and/or CXR assume a supervised classification strategy to detect COVID-19 cases \cite{luz2020towards,oh2020deep,wang2020covid,zhang2020viral}. 
The premise of these supervised classification studies is that sufficient cases of COVID-19 have already occurred, which implies that 
more data is available for training better classifiers. However, it is non-compassionate and catastrophic to let millions of people get infected (and possibly die) in order to collect sufficient data to train classifiers to detect the disease. Furthermore, this approach does not help detect early cases of a pandemic, such as COVID-19, which is necessary to contain and prevent further loss of human life.
We argue that supervised classification for detecting early cases of a pandemic could be regarded as a good strategy \textit{only after}, but not \textit{before} the fact. Therefore, there is a need to postulate alternative problem formulations for detecting early cases of pandemics in order to prevent it from spreading and becoming hard to manage.
    
Anomaly detection \cite{khan2014one} framework offers an alternative paradigm to learn classifiers from only normal samples and identify the not-normal samples as anomaly. In this case, the samples for normal class are readily available (e.g. healthy CXR) at no additional cost; however, the samples for not-normal class (e.g., COVID-19 cases) may be either unavailable, poorly sampled, or too costly to collect (both in terms of dollars and human life). We formulate early detection of COVID-19 cases from CXR as an anomaly detection problem, given the challenge in collecting data from infected persons and the accompanying cost. To demonstrate the validity of our proposal, we present several unsupervised deep learning based anomaly detection approaches that use autoencoders and their adversarial counterparts.
Our results show encouraging performance and opens up a new direction of research for emergency preparedness for future waves of COVID-19 or other potential pandemics.

\section{Related Work}
\label{sec:relatedwork}

Recent breakthroughs in deep learning have been successfully translated into the screening of COVID-19 cases through radiology imaging \cite{luz2020towards,oh2020deep,wang2020covid,ucar2020covidiagnosis}. Most of these works have been conducted in a supervised classification setting. 
As discussed in the previous section, and due to the scope of this paper, we will only discuss those studies that use either one-class classification (OCC) or anomaly detection setup for detecting COVID-19 using CXR.

\citet{zhang2020viral} propose a confidence-aware anomaly detection (CAAD) model that works by a convolutional feature detector model feeding into an anomaly detection module and a confidence prediction module that work together to classify instances of healthy control, viral-pneumonia, and non-viral pneumonia. 
However, CAAD relies on its confidence prediction module and needs positive viral-pneumonia CXR training examples to increase its AUC from $0.836$ to $0.874$. 
\citet{tuluptceva2020anomaly} propose a deep autoencoder with progressively growing blocks with residual connections to detect abnormalities in CXR images. In their model, an autoencoder learns the characteristics of `normal' CXR and detect out-of-distribution anomalous CXR images. The model, however, uses both normal and abnormal cases during the training. The authors use the model for the more general task of detecting abnormalities in CXR images and metastases in the lymph node. 
\citet{li2020robust} present COVID-19 detection from CXR images as a cost-sensitive learning problem to handle high misdiagnosis cost of COVID-19 cases due to their visual similarity to other pneumonia cases. 
This approach assumes availability of sufficient data for COVID-19 cases, which may not be the case during early stages of a pandemic.

It is clear from the literature review that there is a paucity of research using anomaly detection approaches for early identification of pandemic cases.
Research on these techniques can significantly impact the emergency preparedness of public health agencies. Our main contribution is to formulate early detection of pandemic, such as COVID-19, from CXR in an anomaly detection framework and test it on a large publicly available CXR dataset using various unsupervised deep learning approaches. 

\section{Unsupervised Deep Learning based Anomaly Detection}
\label{sec:anomaly}

\begin{figure*}[ht]
    
    \centering
    \includegraphics[width=\textwidth, height=3.5cm]{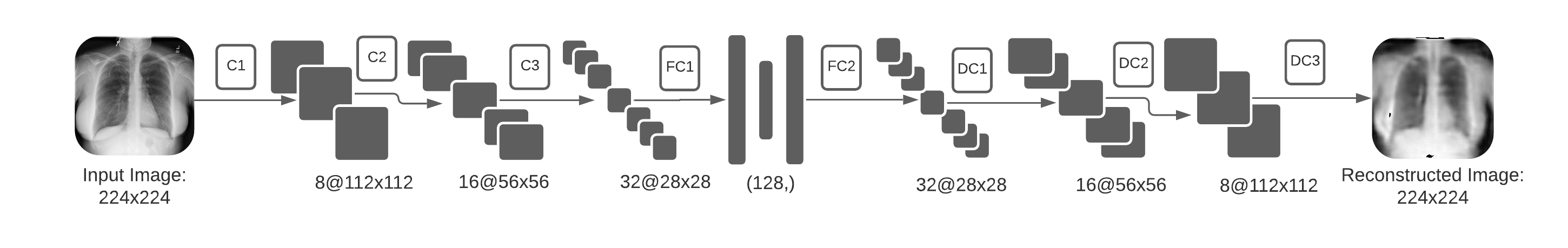}  
    \vspace{-5mm}
    \caption{The convolutional autoencoder architecture for detecting early pandemic cases from CXR}
    \label{fig:architecture}
\end{figure*}

For detecting anomalies in CXR images, we investigate two types of unsupervised deep learning approaches: convolutional autoencoder (CAE)\footnote{The code is available here: https://github.com/faraz2023/COVIDomaly} and  adversarially trained CAE. 
The architecture of the convolutional autoencoder is shown in Figure \ref{fig:architecture}.
The encoder in CAE is composed of three convolution blocks (C1, C2, C3) and a fully connected block (FC1). Each of C1, C2, C3 consists of four layers: a batch normalization layer, a convolutional layer, a leaky ReLU layer, and a Maxpool ($size=2\times2$) layer. The kernel size is $7\times7$, $stride=1$, and $padding=2$, for each convolution layer. C1, C2, and C3 produce outputs consisting of $8$, $16$, and $32$ channels respectively. The fully connected block (FC1) consists of a flattened layer (output size: $25088 \times 1$), and two fully connected layers with leaky ReLU layers, to reduce the dimensiononality of the data to $128$. The decoder replicates the encoder in a reverse manner and 
reconstructs the input image using the latent encoded representation. 
The CAE is only provided with normal cases during the training phase to enable it 
learn features that are good for reconstructing the images from healthy individuals. 
In the test set, the CAE is expected to produce a low reconstruction error for normal cases, and  higher reconstruction error for the anomalous (COVID-19) samples. Therefore, the CAE's reconstruction error is used as a score to detect anomalies.

\vspace{-2mm}
For the adversarial approaches, we employed a CAE and convolutional neural network (CNN) to work in tandem and reconstruct CXR images under two different training frameworks -- (a) a Vanilla adversarial approach, and (b) a Wasserstein adversarial approach. The two approaches had largely the same neural network architectures, but differed in terms of the way  the networks were trained together, which is 
discussed 
below. 


The CAE in all the three experiments remained the same and consisted of an encoder-decoder pair as described above.  Firstly, the input CXR images are resized to $224\times224$ and the number of channels is reduced to $1$. Therefore, the input to the encoder is a gray-scale CXR image; after three (alternate) Convolutions, Maxpooling, Batch Normalization layers, followed by a fully connected layer, the encoder produces a latent representation of dimension $128$. 
The input to the decoder is this latent representation followed by one fully connected layer and three (alternate) Batch Normalization, Convolution Transpose and Upsampling layers to reconstruct an image with the same dimensions as the input CXR.

The Vanilla and Wasserstein adversarial approaches also include a CNN discriminator which is the same for the two experiments with the following differences: For the Vanilla approach the last layer of the CNN is Sigmoid activated, while for the Wasserstein experiment, the last layer of the CNN discriminator output does not have any activation function. 

The CNN discriminator mirrors the architecture of the encoder part of the CAE for the most part. It is composed of three convolution blocks (C1, C2, C3) and a fully connected block (FC1). Each of C1, C2, C3 consists of four layers: a batch normalization layer, a convolutional layer, a  ReLU layer, and a Maxpool (size= $2 \times 2$) layer. The kernel size is $7 \times 7$, stride= $1$, and padding = $2$, for each convolutional layer. C1, C2, and C3 produce outputs consisting of $8$, $16$, and $32$ channels respectively. The fully connected block (FC1) consists of a flattened layer (output size: $25088 \times 1$), and two fully connected layers with ReLU layers, to reduce the dimensionality of the data  first to $256$, then to $128$, and finally to $1$. In the Vanilla adversarial experiment, this scalar is then fed to a Sigmoid function, while in the Wasserstein adversarial experiment the scalar itself is the output of the discriminator. 

These networks are trained on two settings of `normal' cases: (i) CXR from healthy persons only, and (ii) CXR from healthy and  non-COVID pneumonia patients. Both healthy and/or non-COVID pneumonia cases are considered as ``normal'' classes because abundant data is readily available for them. During testing, the data for both normal and COVID-19 CXR is presented to the autoencoders. 

\section{Dataset}
\label{dataset}

To test above-described unsupervised deep learning approaches for detecting early cases in COVID-19 pandemic, we use COVIDx dataset \cite{wang2020covid}, which is an ongoing project, i.e., the number of available examples may change over time. The snapshot of COVIDx used in this paper consists of $8851$ CXR images from ``healthy'' individuals, $6052$ CXR images of ``non-COVID pneumonia'', and $498$ CXR images of ``COVID-19'' infected persons.

\section{Experimental Results}
\label{experimental_results}

\subsection{Evaluation}
\label{evaluation}

For both the settings (i) and (ii) described above in Section \ref{sec:anomaly}, 3-fold cross-validation is performed; for each fold the network is trained for $750$ epochs with a batch-size of $100$. 
In each fold of the $3$-fold cross validation for setting (i),  the autoencoders was  trained on $2/3$ of the normal cases 
and tested on the remaining $1/3$ of the normal cases
and $1/3$  COVID-19 cases.  
For each fold for setting (ii),  the autoencoders were trained on $2/3$ of the normal plus non-COVID-19 pneumonia cases
and tested on the  remaining $1/3$ of the normal plus $1/3$ of non-COVID-19 pneumonia cases 
and $1/3$  COVID-19 cases. 
The process is repeated three times. The performance metric used to evaluate the models for both the settings is Area Under the Curve (AUC) of the Receiver Operating Characteristic (ROC) plot. We calculated AUC using two methods. Firstly, we calculated AUC for each fold and reported the mean ($AUC_\mu$) and standard deviation ($AUC_\sigma$) across 3-folds. Secondly, we calculated pooled AUC ($AUC_p$) across 3-folds by concatenating the reconstruction error after each fold and calculating the AUC after the completing the 3-fold cross validation.

For the autoencoder approach, the loss function is Mean Squared Error (MSE), with Adam optimizer 
with an initial learning rate of $10^{-3}$, which is then progressively reduced by $3 \times 10^{-4}$ after every $250$ epochs. 
For the Vanilla Adversarial approach, the networks are trained for $750$ epochs with a batch size of $128$. The loss function for the CAE is the average MSE computed over a training batch. The loss for the discriminator is also the regular vanilla GAN discriminator loss. Each neural network utilizes Adam optimizer with a learning rate of $10^{-3}$.
For the Wasserstein Adversarial approach the networks are trained for $750$ epochs with a batch size of $128$. The loss function for the CAE is the sum of MSE and Wasserstein Loss. The loss for the discriminator is also the Wasserstein GAN discriminator loss. Each neural network utilizes RMSProp optimizer with a learning rate of $15 * 10^{-5}$. Following the  Wasserstein GAN approach, we train the discriminator more as compared to the CAE ($5$ times more), and its weights are clipped to the range of $[-0.01, 0.01]$.

As shown in Table \ref{tab:AUC_table}, using CAE, for case (i), we obtained $AUC_{\mu}=0.765$,  $AUC_{\sigma}=0.023$, and $AUC_p=0.765$. For case (ii), we obtained $AUC_{\mu}=0.690$,  $AUC_{\sigma}=0.020$, and $AUC_p=0.690$. These results suggest that COVID-19 can be detected as an anomaly when the CAE is trained only on healthy CXR images without seeing COVID-19 cases during training. However, when the normal class comprises of both healthy and non-COVID-19 pneumonia CXRs, the performance deteriorates ($AUC_{\mu}=0.690$, $AUC_{\sigma}=0.020$, and$AUC_p=0.690$). This could be due to the fact that COVID-19 pneumonia CXRs may have high resemblance with non-COVID-19 pneumonia CXRs in terms of radiographic appearances of multifocal ground glass opacities, linear opacities, and consolidation \cite{Cleverleym2426}. 
To verify the similarity between COVID-19 and non-COVID-19 pneumonia cases, we trained CAE on only CXR of non-COVID-19 pneumonia and tested on CXR of non-COVID-19 pneumonia and COVID-19 in a 3-fold cross-validation manner described above. For this setting, we obtained $AUC_{\mu}=0.678$,  $AUC_{\sigma}=0.018$, and $AUC_p=0.677$, which is very similar in performance to setting (ii). This shows that CAE is still able to detect COVID-19 cases with reasonable confidence even with only seeing non-COVID-19 pneumonia CXR during training the model.

\begin{table}[htbp]
  \centering
  \caption{AUC of ROC using the CAE}%
  \label{tab:AUC_table}
    \begin{tabular}{|p{35mm}|l|l||l|}
        \hline
        \multirow{2}*{\textbf{Training Cases}} & \multicolumn{3}{c|}{\textbf{Performance}} \\ \cline{2-4} 
        & $AUC_\mu$ & $AUC_\sigma$ & $AUC_p$\\ \hline
        (i) Healthy & \textbf{0.765} & 0.023 & \textbf{0.765}\\ \hline
        (ii) Healhty + Pneumonia & 0.690 & 0.020 & 0.690\\ \hline
        (iii) Pneumonia & 0.678 & 0.018 & 0.677\\ \hline
    \end{tabular}
\end{table}

Since setting (i) gave better results, we trained both the adversarial approaches under setting (i). Table \ref{tab:AUC_adver_table} shows that with Vanilla approach, $AUC_{\mu}=0.755,  AUC_{\sigma}=0.017$, and $AUC_p=0.762$; with Wasserstein trained CAE, $AUC_{\mu}=0.511,  AUC_{\sigma}=0.089$, and $AUC_p=0.537$. The adversarial trained CAE does not show improvement over regular CAE.

\begin{table}[htbp]
  \centering
  \caption{AUC of ROC using adversarial CAE}%
  \label{tab:AUC_adver_table}
    \begin{tabular}{|p{25mm}|l|l|l||l|}
        \hline
        \multirow{2}*{\textbf{Experiment}} &
        \multirow{2}*{\textbf{Training Cases}} & \multicolumn{3}{c|}{\textbf{Performance}} \\ \cline{3-5} 
         & & $AUC_\mu$ & $AUC_\sigma$ & $AUC_p$\\ \hline
        
        (i) Vanilla & Healthy & 0.755 & 0.017 & 0.762\\ \hline
        (ii) Wasserstein & Healhty & 0.511 &  0.089 & 0.537\\ \hline

    \end{tabular}
\end{table}

\section{Conclusions and Future Work}
\label{conclusions_future_work}

In this paper, we presented an unsupervised deep learning approach to detect early cases of a pandemic, such as COVID-19. 
Although COVID-19 has already turned into a pandemic, we argue that using anomaly detection models, early cases in a future pandemic or newer waves of an existing pandemic could be detected with high confidence by only utilizing the abundantly available normal CXR images across various health organizations.
From the deep learning perspective, this preliminary work would benefit from incorporating feature interpretability \cite{jetley2018learn,sundararajan2017axiomatic} and combination of multiple loss functions, such as gradient loss and intensity loss that have shown good performance in other domains. 

\thispagestyle{plain}
\printbibliography


\end{document}